\documentclass{article}
\usepackage{graphics}
 \usepackage{graphicx}

 \usepackage{epsfig}

\usepackage{amssymb}
\usepackage{amsmath}
 \usepackage{amsthm}

\begin{document}

\title{Phonon-assisted thermoelectric effects in a two-level molecule}

\author{M.˜ Bagheri Tagani,
        H.˜ Rahimpour Soleimani,\\
        \small{Department of physics, University of Guilan, P.O.Box 41335-1914, Rasht, Iran}}

\maketitle

\begin{abstract}
Thermoelectric properties of a two-level molecule attached to the
metallic electrodes are analyzed using the equation of motion
technique within the  Green's function formalism. Results show
that the electrical conductance is strongly dependent on the
electron and phonon temperatures and the electron-phonon coupling
strength. In addition, it is observed that the thermal
conductance peaks in the electron-hole symmetry points are
vanished in the presence of the strong electron-phonon
interaction. It is also found that the figure of merit is
strongly suppressed in the strong electron-phonon interaction.
The violation of the Wiedemann-Franz law is also observed coming
from the Coulomb interactions.
\end{abstract}

\section{Introduction}
\label{Introduction} Transport phenomena through molecular
devices have been attracted a lot of attention during recent two
decades.  Experimental and theoretical results show that the
electron transport through the molecular transistors results in
novel and interesting effects such as Coulomb and spin blockade
effects~\cite{ref1,ref2,ref3,ref4}, Kondo
effect~\cite{ref5,ref6,ref7}, negative differential
conductance~\cite{ref8,ref9,ref10}, and so on. In addition, it has
been predicted that the molecular devices can be efficient for
conversion of heat into electric energy~\cite{ref11,ref12}.
Thermoelectric efficiency of a device is indicated by a
dimensionless quantity called figure of merit,
$ZT=S^2G_eT/\kappa$, where $G_e$ is the electrical conductance
and $S$ is the thermopower. $\kappa$ is the thermal conductance
which is the sum of the electronic contribution, $\kappa_e$, and
the lattice thermal conductance, $\kappa_p$, and $T$ denotes the
operating temperature. In bulky samples and according to the
Wiedemann-Franz Law, the ratio of the electrical conductance to
the thermal conductance is related to the operating temperature.
Research shows that the strong Coulomb repulsions, discreteness
of energy levels, interference effects and etc. in nanostructures
result in the violation of the Wiedemann-Franz
law~\cite{ref13,ref14,ref15} and as a consequence, the increase of
$ZT$. For these reasons, the investigation of the thermoelectric
properties of the systems composed of a single or two quantum dots
has attracted a lot of attention in recent
years~\cite{ref15,ref16,ref17,ref18,ref19,ref20,ref21,ref22,ref23,ref24,ref25}.

\par Electron-phonon interaction (EPI) is an interesting and
important phenomenon in the molecular devices which can
significantly affect the transport characteristics of the
devices. The center of mass oscillation of the
molecule~\cite{ref26}, or thermally-induced acoustic
phonons~\cite{ref27} can be the origin of the coupling between the
electronic degree of freedom and the vibrational degree of
freedom. The influence of the EPI on the electronic
characteristics of the molecular transistors and the carbon
nanotube quantum dots has been extensively studied using both rate
equation approach~\cite{ref28,ref29,ref30,ref31,ref32} and the
 Green's function
formalism~\cite{ref33,ref34,ref35,ref36,new}. However, the
influence of the EPI on the thermoelectric properties of the
molecular devices is an interesting issue needing more attention.
Koch and co-workers~\cite{ref37} analyzed the thermopower in a
single level molecule by means of the master equation.
Furthermore, the effect of the EPI on the thermoelectric
properties of a single-level quantum dot was studied in a few
articles using the  Green's function
formalism~\cite{ref38,ref39,ref40}.
\par In this article, we consider a two-level
molecule in which the electronic degree of freedom is coupled to
the vibrational degree of freedom. Using the equation of motion
technique within the  Green's function formalism and using the
polaronic transformation, the density of the states of the
molecule (DOS) and the thermoelectric properties of the device
are analyzed. With respect to the fact that the phonon subsystem
has a smaller heat capacity than the electron subsystem, it is
completely probable an electron-phonon nonequilibrium
thermodynamics is dominant so that the electron
temperature,$T_e$, and phonon temperature, $T_p$, are different.
The difference can result in the appearance of the phonon
absorption sidebands in the DOS in the low electron temperatures
which are not seen in the EP thermal equilibrium. The effects of
the EP nonequilibrium and EP coupling strength on the
thermoelectric properties of the system are analyzed in detail.
In the next section, formalism is presented. Section 3 is devoted
to the numerical results and in the end; some sentences are given
as a summary.

 \section{Model and formalism}
 \label{Model}
We consider a two-level molecule coupled to the normal metal
electrodes. The Hamiltonian describing the system is given as
\begin{align}\label{Eq.1}
H&=\sum_{\alpha k\sigma}\varepsilon_{\alpha
k\sigma}c^{\dag}_{\alpha k\sigma}c_{\alpha
k\sigma}+\sum_{i=1,2\sigma}\varepsilon_{i\sigma}n_{i\sigma}+\frac{1}{2}\sum_{ij\sigma\sigma'}U_{ij}n_{i\sigma}n_{j\sigma'}\\\nonumber
&\omega a^{\dag}a+\sum_{i\sigma}\lambda
n_{i\sigma}[a^{\dag}+a]+\sum_{\alpha k\sigma i}[V_{\alpha
k\sigma}^{i}c^{\dag}_{\alpha k \sigma}d_{i\sigma}+H.C.],
\end{align}
where $c^{\dag}_{\alpha k\sigma}$ creates an electron with wave
vector $k$, spin $\sigma$, and energy $\varepsilon_{\alpha
k\sigma}$ in lead $\alpha$. $d_{i\sigma}^{\dag}$ creates an
electron in the $i$th level of the molecule with energy
$\varepsilon_{i\sigma}$, while $U_{ii}$ stands for the on-site
Coulomb repulsion and $U_{ij}$ ($i\neq j$) denotes the inter-level
Coulomb repulsion. The energy levels of the molecule are tuned by
a gate voltage thus we set
$\varepsilon_{i\sigma}=\varepsilon_{i\sigma}^{0}+V_g$ so that
$\varepsilon^2_0=\varepsilon^0_1+\Delta$ where $\Delta$ denotes
the level spacing. $a^{\dag}(a)$ is the creation (annihilation)
operator for the phonons with energy $\omega$ and $\lambda$
describes the electron-phonon coupling strength. The last term in
the above equation describes the tunneling process between the
electrodes and the molecule and $V_{\alpha k\sigma}^{i}$ is the
coupling strength between the lead $\alpha$ and the $i$th energy
level. $V_{\alpha k\sigma}^{i}$ induces a tunneling rate from the
molecule to the electrode $\alpha$,
$\Gamma^{i}_{\alpha\sigma}=2\pi\sum_{k\in
\alpha}\rho_{\alpha}|V_{\alpha k\sigma}^{i}|^2$ where
$\rho_{\alpha}$ is the electronic density of the lead $\alpha$.
\par The EPI can be eliminated using polaronic transformation~\cite{ref41},
$\tilde{H}=e^{S}He^{-s}$, where
$S=\textrm{exp}(\sum_{i\sigma}\lambda n_{i\sigma}[a^{\dag}-a])$.
Thus, equation~\eqref{Eq.1} becomes
\begin{align}\label{Eq.2}
\tilde{H}&=\sum_{\alpha k\sigma}\varepsilon_{\alpha
k\sigma}c^{\dag}_{\alpha k\sigma}c_{\alpha
k\sigma}+\sum_{i\sigma}\tilde{\varepsilon}_{i\sigma}n_{i\sigma}+\frac{1}{2}\sum_{ij\sigma\sigma'}\tilde{U}_{ij}n_{i\sigma}n_{j\sigma}\\\nonumber
&\omega a^{\dag}a+\sum_{\alpha k\sigma i}[\tilde{V}_{\alpha
k\sigma}^{i}c^{\dag}_{\alpha k\sigma}d_{\sigma}+H.C],
\end{align}
where
$\tilde{\varepsilon}_{i\sigma}=\varepsilon_{i\sigma}-\lambda^2/\omega$,
and $\tilde{U}_{ij}=U_{ij}-2\lambda^2/\omega$. The polaronic
transformation results in the renormalization of the molecule
levels and Coulomb repulsions. In addition, the tunneling
amplitude is transformed into $\tilde{V}_{\alpha
k\sigma}^{i}=V_{\alpha k\sigma}^{i}X$ where
$X=exp(-\frac{\lambda}{\omega}[a^{\dag}-a])$ is the phonon
operator. In the following, we replace the operator $X$ with its
expectation value,
$<X>=exp(-\frac{\lambda}{\omega}(N_p+1/2))$~\cite{ref41} where
$N_p$ stands for the averaged number of phonons. This
approximation is used in the localized polaron and is valid when
the tunneling amplitude is smaller than the EPI, i.e. $V_{\alpha
k\sigma}<<\lambda$. Hence, the tunneling rate is renormalized
according to $\tilde{\Gamma}^{i}_{\alpha
\sigma}=\Gamma^{i}_{\alpha\sigma}<X>^2$. $<X>^2$ just narrows the
tunneling rate.
\par In order to compute the electronic current, ${I_e}$, and the
heat flux, ${I_Q}$, the  Green function formalism is used so that
the electronic current and heat current are expressed
as~\cite{ref42,ref43}
\begin{subequations}\label{Eq.3}
\begin{align}
  I_e&=\frac{-e}{h}\sum_{i\sigma}\int_{-\infty}^{\infty}d\varepsilon\frac{\Gamma^{i}_{L\sigma}\Gamma^{i}_{R\sigma}}{\Gamma^{i}_{L\sigma}+\Gamma^{i}_{R\sigma}}
  [f_{L}(\varepsilon)-f_{R}(\varepsilon)]\textit{{Im}} G^{r}_{i\sigma}(\varepsilon)\\
  I_Q&=\frac{-1}{h}\sum_{i\sigma}\int_{-\infty}^{\infty}d\varepsilon
  \frac{\Gamma^{i}_{L\sigma}\Gamma^{i}_{R\sigma}}{\Gamma^{i}_{L\sigma}+\Gamma^{i}_{R\sigma}}(\varepsilon-\mu)
  [f_{L}(\varepsilon)-f_{R}(\varepsilon)]\textit{Im} G^{r}_{i\sigma}(\varepsilon)
  \end{align}
\end{subequations}
where
$f_{\alpha}(\varepsilon)=(1+\textrm{exp}((\varepsilon-\mu_\alpha))/kT_e)^{-1}$
is the Fermi distribution function and $\mu_\alpha$, and $T_e$ are
the chemical potential and electron temperature in the lead
$\alpha$. It is worth noting that equation~\eqref{Eq.3} is
computed with respect to the tunneling part of the Hamiltonian,
equation~\eqref{Eq.1}, so the bare tunneling rates are used.
However, the influence of the EPI on the electronic and heat
currents is saved in the Green function, $G_{i\sigma}^{r}$.
-$Im{G^{r}_{i\sigma}(\varepsilon)}$ is the DOS of the level $i$
composed of the phononic and electronic parts which can be
computed by evaluating the phonon part of the trace by Feynman
disentangling technique~\cite{ref41}, and using the Keldysh
equations~\cite{ref43} for the lesser and greater Green
functions, as follows~\cite{ref33}
\begin{align}\label{Eq.4}
  \textit{Im} G^{r}_{i\sigma}(\varepsilon)&=\sum_{\alpha=L,R}\sum_{n=-\infty}^{\infty}\frac{L_n}{\tilde{\Gamma}^{i}_{L\sigma}+\tilde{\Gamma}^{i}_{R\sigma}}
  [\tilde{\Gamma}^{i}_{\alpha\sigma}f_{\alpha}(\varepsilon+n\omega)\textit{Im} \tilde{G}^{r}_{i\sigma}(\varepsilon+n\omega)\\\nonumber
  &+\tilde{\Gamma}^{i}_{\alpha\sigma}[1-f_{\alpha}(\varepsilon-n\omega)]\textit{Im} \tilde{G}^{r}_{i\sigma}(\varepsilon-n\omega)],
\end{align}
where $L_n$, for nonzero temperatures, is given as
\begin{equation}\label{Eq.5}
  L_n=e^{-(\lambda/\omega)^2(1+2N_{ph})}(\frac{N_{ph}+1}{N_{ph}})^{n/2}I_n(2(\lambda/\omega)^2\sqrt{N_{ph}(N_{ph+1})})
\end{equation}
where $I_n(x)$ are the Bessel functions of the complex argument,
and $N_{ph}=(\textrm{exp}(\omega/kT_{p})-1)^{-1}$ is the averaged
number of phonons and $T_p$ stands for the phonon temperature.
Indeed, the thermal equilibrium between the electrons in the
electrodes and phonons in the molecule can be broken due to
smaller heat capacity of the phonons.
$\tilde{G}^{r}_{i\sigma}(\varepsilon)$ is the dressed retarded
Green function governed by $\tilde{H}$. To compute
$\tilde{G}^{r}_{i\sigma}$, we follow the procedure introduced by
Chang and co-workers~\cite{ref44}. Therefore,
$\tilde{G}^{r}_{i\sigma}$ is expressed as
\begin{equation}\label{Eq.6}
  \tilde{G}^{r}_{i\sigma}(\varepsilon)=\sum_{k=0}^{2}p_k(\frac{1-N_{i-\sigma}}{E_{i\sigma}-\Pi_{k}}+\frac{N_{i\sigma}}{E_{i\sigma}-U_{ii}-\Pi_{k}})
\end{equation}
where
$E_{i\sigma}=\varepsilon-\varepsilon_{i\sigma}+1/2(\tilde{\Gamma}^{i}_{L\sigma}+\tilde{\Gamma}^{i}_{R\sigma})$,
and, the summation is over all possible configurations in which
the level j ($j\neq i$) is empty,
$p_0=1+<n_{j\sigma}n_{j-\sigma}>-(N_{j\sigma}+N_{j-\sigma})$,
singly, $p_{1}=N_{j\sigma}+N_{j-\sigma}$, or doubly,
$p_{2}=<n_{j\sigma}n_{j-\sigma}>$, occupied. $p_k$ denotes the
probability factor of a configuration expressed by one-particle
$N_{i\sigma}=<d^{\dag}_{i\sigma}d_{i\sigma}>_{\tilde{H}}$ and
two-particle $<n_{i\sigma}n_{i-\sigma}>_{\tilde{H}}$ occupation
numbers. $\Pi_{k}$ denotes the sum of Coulomb repulsions seen by
an electron in level $i$ due to other electrons in configuration
$k$, in which the level $j (j\neq i)$ is empty $\Pi_0=0$, singly,
$\Pi_{1}=U_{12}$, or doubly, $\Pi_{2}=2U_{12}$ occupied.
\par In the linear response theory for the electronic current and
heat flux (equation~\eqref{Eq.3}), the electrical conductance is
equal to $G_e=-\frac{e^2}{T_e}L_{11}$, thermopower is
$S=-\frac{1}{eT_e}\frac{L_{12}}{L_{11}}$, and the thermal
conductance is given by
$\kappa_e=\frac{1}{T_e^2}[L_{22}-\frac{L_{12}^2}{L_{11}}]$. The
linear response coefficients are given by
\begin{subequations}\label{Eq.7}~\cite{ref45}
  \begin{align}
  L_{11}&=\frac{T_e}{h}\sum_{i\sigma}\int d\varepsilon
  \frac{\Gamma^{i}_{L\sigma}\Gamma^{i}_{R\sigma}}{\Gamma^{i}_{L\sigma}+\Gamma^{i}_{R\sigma}}
  Im G^{r}_{i\sigma}(\varepsilon)
  (-\frac{\partial{f(\varepsilon)}}{\partial{\varepsilon}})_{T_e}\\
  L_{12}&=\frac{T_e^2}{h}\sum_{i\sigma}\int d\varepsilon
  \frac{\Gamma^{i}_{L\sigma}\Gamma^{i}_{R\sigma}}{\Gamma^{i}_{L\sigma}+\Gamma^{i}_{R\sigma}}
  Im G^{r}_{i\sigma}(\varepsilon)
  (\frac{\partial{f(\varepsilon)}}{\partial{T_e}})_{\mu}\\
  L_{22}&=\frac{T_e^2}{h}\sum_{i\sigma}\int d\varepsilon
  \frac{\Gamma^{i}_{L\sigma}\Gamma^{i}_{R\sigma}}{\Gamma^{i}_{L\sigma}+\Gamma^{i}_{R\sigma}}
  (\varepsilon-\mu) Im G^{r}_{i\sigma}(\varepsilon)
  (\frac{\partial{f(\varepsilon)}}{\partial{T_e}})_{\mu}
  \end{align}
\end{subequations}

\par For simulation purposes, we use half band width, $D$, as
energy unit and set $\omega=D/50$, and $U_{ii}=2U_{12}=D/10$. We
also take the renormalized tunneling rate,
$\tilde{\Gamma}^{i}_{\alpha\sigma}$ as an input parameter and
ignore the narrowing the width of the tunneling rate due to the
EPI and assume the wave functions of the molecular levels are
identical, i.e.
$\Gamma^{1}_{\alpha\sigma}=\Gamma^{2}_{\alpha\sigma}$ . It has
been shown that the EPI does not narrow the width of tunneling
rate in some cases~\cite{ref46,ref47}.

\begin{figure}[htb]
\begin{center}
\includegraphics[height=100mm,width=100mm,angle=0]{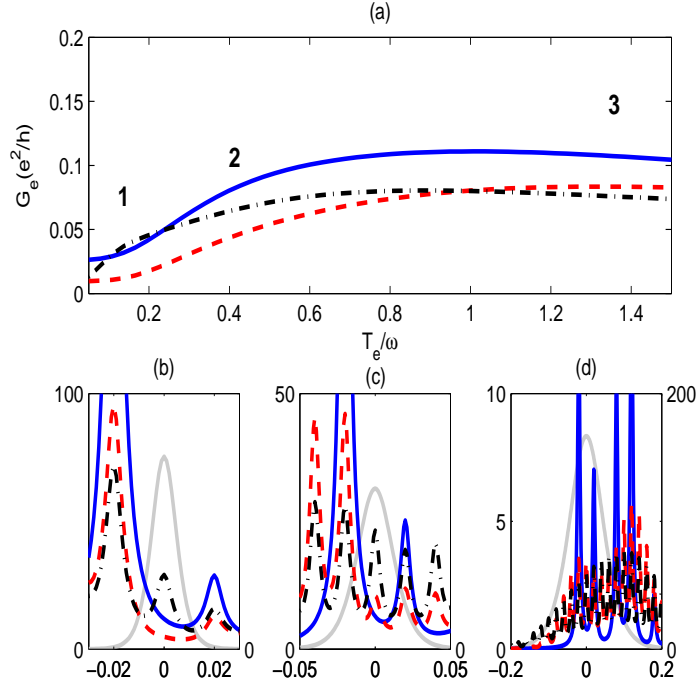}
\caption{(a) Electrical conductance as a function of electron
temperature for $\lambda=0$ (solid line), $\lambda=1,T_p=T_e$
(dashed line), and $T_p=5T_e$ (dash-dotted line). (b)-(d) show DOS
in the parts1-3, respectively.
$-\frac{\partial{f(\varepsilon)}}{\partial{\varepsilon}}$ is
plotted in gray. $\Gamma^{i}_{\alpha\sigma}=0.2\omega$, and
$\Delta=2\omega$ is the level spacing.}\label{fig:1}       % Give a unique label
\end{center}
\end{figure}

\section{Results and discussions }
\label{Numerical results} Figure 1a describes the electrical
conductance, $G_e$, as a function of temperature in the absence
and presence of the electron-phonon interaction. Results show
that $G_e$ is uniformly increased by increase of temperature.
However, behavior of it is strongly dependent on the electron and
phonon temperatures and EPI. One can divide figure 1a to three
parts. In part 1, low temperatures, $kT_e<0.2\omega$, the
electrical conductance resulting from the EP nonequilibrium is
more. Figure 1b can help us understand the reason. In low
temperatures the DOS of the molecule exhibits a vibrational
absorption sideband (dash-dotted line) located near the chemical
potential of the leads. Such peak is not observed in the elastic
transport (solid line) or in the presence of the thermal
equilibrium between electrons and phonons (dashed line). Note
that there are no phonons in the molecule when phonons and
electrons are in the thermal equilibrium because of
$kT_p<<\omega$. In region 2, mediated temperatures,
$0.2<kT_e<\omega$, the electrical conductance of the elastic
transport is more. However, EP nonequilibrium can enhance $G_e$ in
comparison with the EP equilibrium. It is worth noting that $G_e$
resulted from the EPI is independent of the temperature
difference between electrons and phonons for $kT_e=\omega$. In
region 3, high temperatures, $kT_e>\omega$, $G_e$ approaches
saturated values. The DOS of the molecule and $(-f'(\varepsilon))$
are plotted in figures 1a-d for parts 1-3, respectively.
$f'(\varepsilon)$ is a Lorentzian function whose center is located
in $\mu$ and its width depends on the electron temperature.
Therefore, $f'(\varepsilon)$ becomes wider with increase of the
temperature so that more part of the DOS involves in the electron
transport.
\begin{figure}[htb]
\begin{center}
\includegraphics[height=100mm,width=110mm,angle=0]{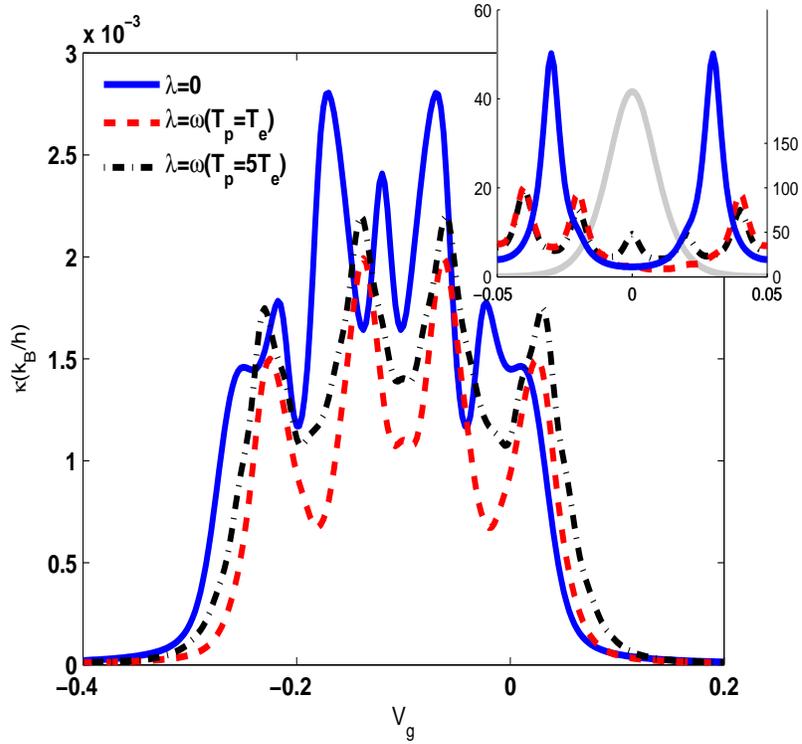}
\caption{Thermal conductance as a function of gate voltage.
Inset shows the DOS when $V_g=-1/2(\Delta+\tilde{U_1}+2\tilde{U_{12}})$. $T_e=0.3\omega$}\label{fig:2}       % Give a unique label
\end{center}
\end{figure}
\par The electrical conductance, $\kappa_e$, is plotted
in figure 2 as a function of the gate voltage. In $\lambda=0$, the
electrical conductance has seven peaks. Four peaks are located in
resonance energies i.e. $V_g=\varepsilon_1^0$,
$V_g=-(\Delta+\tilde{U}_{12})$,
$V_g=-(\varepsilon_1^0+\tilde{U}_1+\tilde{U}_{12})$, and
$V_g=-(\Delta+\tilde{U}_2+2\tilde{U}_{12})$, in which the
electrical conductance is maximum because electrons can easily
tunnel to the molecule. Other peaks are located in electron-hole
symmetry points, i.e. $V_g=-1/2(\Delta+\tilde{U}_{12})$,
$V_g=-1/2(\Delta+\tilde{U}_1+2\tilde{U}_{12})$, and
$V_g=-1/2(\Delta+\tilde{U}_1+\tilde{U}_2+3\tilde{U}_{12})$. In
these points, thermopower is zero. In the electron-hole symmetry
points, electrons and holes participate in the charge and energy
transport with the same weight. Although they carry the charge in
the opposite directions, they carry the energy in the same
direction resulting in the appearance of the peak in the thermal
conductance spectrum. In addition, it is observed that the height
of the peaks is strongly dependent on the electron population of
the molecule, so that the peaks are lower in the voltages
$V_g=\varepsilon_1^0$, and
$V_g=-(\Delta+\tilde{U}_2+2\tilde{U}_{12})$ which the molecule
is, respectively, approximately empty or fully occupied. The
peaks are higher when the molecule is partially occupied. The
dependence of the DOS on the one- and two-electron populations
gives rise to such behavior. In the presence of the EPI, the
height of the peaks is significantly reduced. Furthermore, peaks
located in the electron-hole symmetry points are not well
observed. Inset shows the DOS in
$V_g=-1/2(\Delta+\tilde{U}_1+2\tilde{U}_{12})$. It is clearly
observed the DOS is completely symmetric around the chemical
potential of the leads in the elastic transport, while the EPI
disturbs the symmetry leading to the disappearing the peaks of
the thermal conductance in the electron-hole
 symmetry points. However, interaction of the EP nonequilibrium can
 slightly increase $\kappa_e$ due to the existence of the phonon absorption side peaks.
\begin{figure}[htb]
\begin{center}
\includegraphics[height=100mm,width=110mm,angle=0]{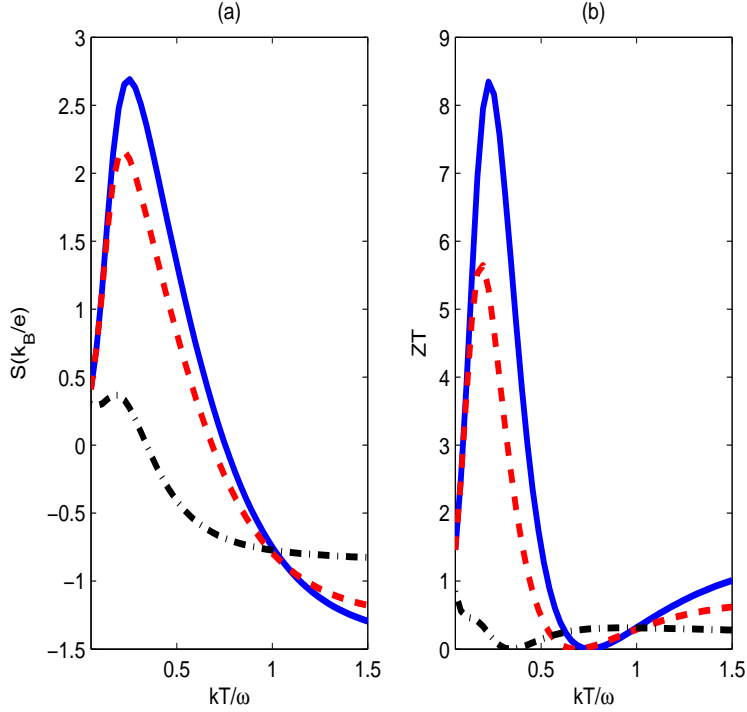}
\caption{(a) Thermopower and (b) figure of merit as a function of
electron temperature. Parameters are
the same as fig.1.}\label{fig:3}       % Give a unique label
\end{center}
\end{figure}
\par Figure 3 describes the behavior of the thermopower and
figure of merit as a function of temperature in the absence and
presence of the EPI. It is observed the thermopower is
increasingly enhanced in the low temperatures, whereas it is
reduced in the high temperatures. Because thermopower has the
main contribution in the figure of merit, $ZT$ follows the same
behavior. Furthermore, results show that the thermopower is
independent of the EPI if $kT_e=\omega$. In low temperatures,
holes participate in the generation of the current because
$\varepsilon_1$ is below the resonance and electrons from the
colder lead (right lead) can tunnel to the warmer one (left
lead). Note that $S$ in the unit of $k/e$ is negative
 because of negative electron charge.  With increase of temperature
  resulting in the excitation of the electrons toward above the
  chemical potential , electrons from the warmer lead can tunnel
  to the colder one using the level $\varepsilon_2$. It is so interesting that
  the sign of the thermopower changes faster in the presence
  of the EPI. It comes from the fact that the electrons and
  holes can more easily tunnel from one lead to the other by
  the phonon absorption processes.  In addition, the magnitude
  of $ZT$ is significantly reduced in the presence of the EPI.
  However, EP nonequilibrium can enhance $ZT$ when $0.6\omega<kT_e<\omega$  because of
  the increase of the thermopower (figure 3a).  $ZT$ approaches
  the lower values in the high temperatures due to the inter-level
  and intra-level Coulomb repulsions (proximity effect).
  Such behavior was analyzed for a triple quantum dot system by Kuo and co-workers~\cite{ref17}.

\begin{figure}[htb]
\begin{center}
\includegraphics[height=80mm,width=80mm,angle=0]{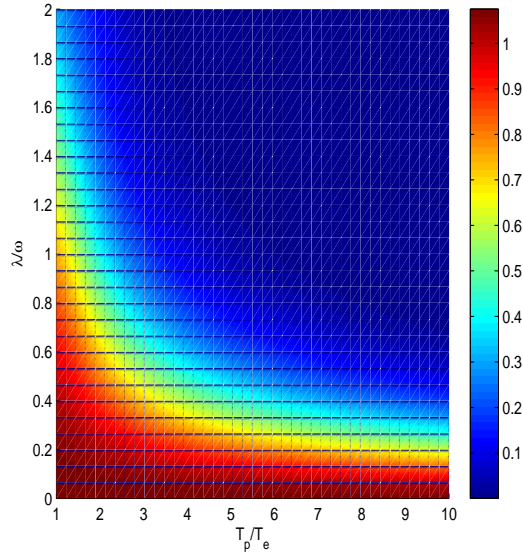}
\caption{ Color map of the figure of merit as a function of
phonon temperature and EP coupling strength. $kT_e=0.2\omega$.}\label{fig:4}       % Give a unique label
\end{center}
\end{figure}
\par The dependence of the figure of merit on the phonon
temperature and EP coupling strength is analyzed in figure 4. In
$\lambda=0$, $ZT$ is independent of the phonon temperature so that
its magnitude is constant and maximum. With increase of $\lambda$
the magnitude of $ZT$ is significantly reduced by increase of
$T_p$. However, in weak $\lambda$s, $\lambda<0.4\omega$, $ZT$ is
nearly constant but in the strong $\lambda$s, $ZT$ is suppressed
because of the reduction of the thermopower and increase of the
thermal conductance.
\begin{figure}[htb]
\begin{center}
\includegraphics[height=80mm,width=80mm,angle=0]{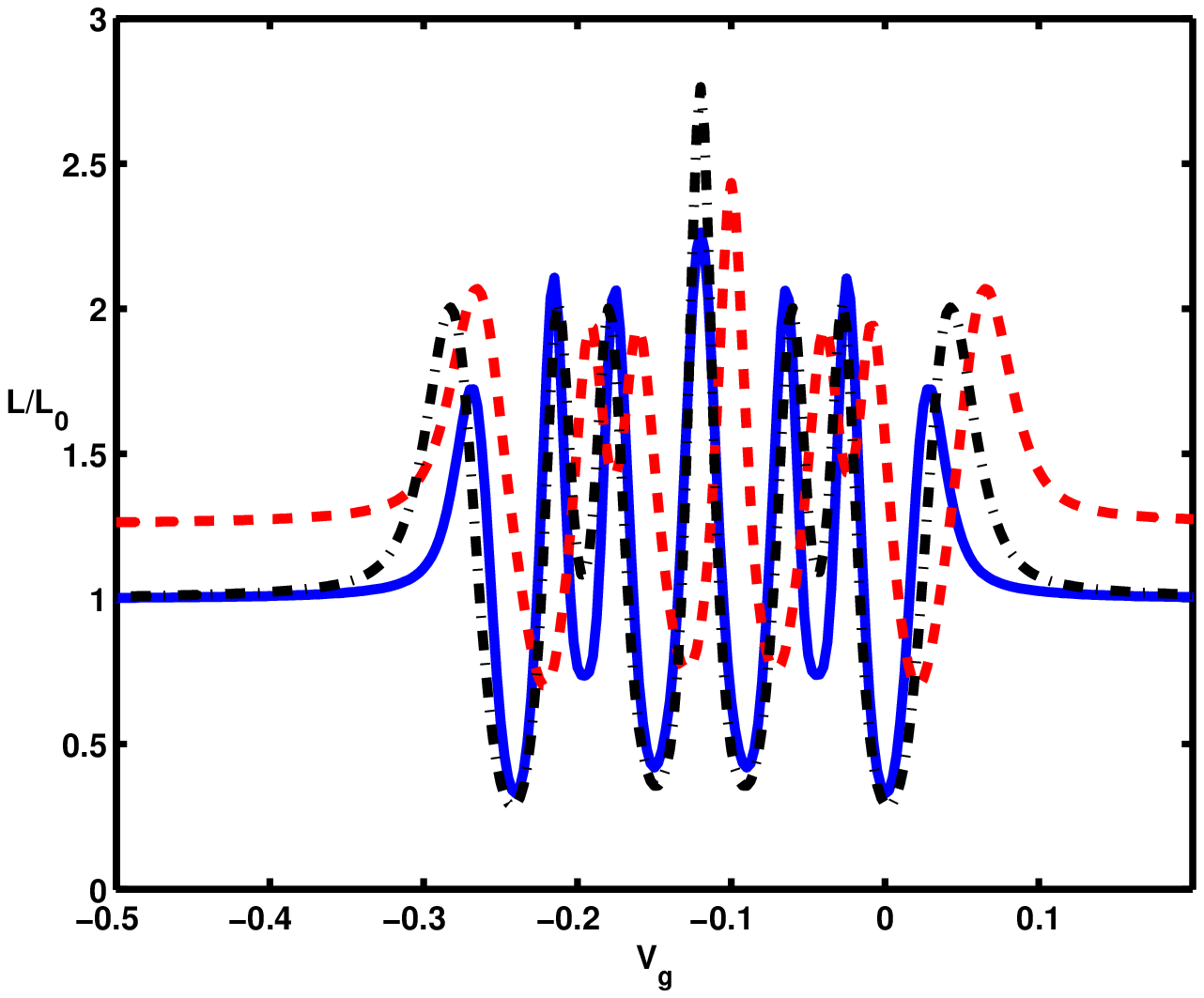}
\caption{Lorentz ratio as a function of gate voltage. Parameters are the same as fig. 2.}\label{fig:5}       % Give a unique label
\end{center}
\end{figure}
\par The Lorentz ration, $L/L_0$, where $L_0=\pi^2k^2/3e^2$ is plotted in figure 5
as a function of the molecule energy levels and in figure 6 as a
function of temperature. The Coulomb interactions give rise to
the violation of the Wiedemann-Franz law. It is observed that the
Lorentz ratio oscillates as a function of the energy level. When
the molecule is completely full or empty, the Lorentz ratio is
one for elastic transport because the transport in this regime is
dominant by higher order tunneling, while it is bigger than unity
for the EP nonequilibrium due to the phonon assisted tunneling.
The Lorentz ratio is significantly increased in the Coulomb
blockade regime. On the other hand, the dependence of the DOS on
the one- and two- electron populations leads to the asymmetry of
the Lorentz ratio. The EP nonequilibrium increases the Lorentz
ratio near the chemical potential of the leads. The behavior of
the Lorentz ratio as a function of temperature is different in
the absence and presence of the EPI. In the presence of the EP
nonequilibrium, the Lorentz ratio is lesser than one because the
electrical conductance is increased due to the phonon
absorption-assisted tunneling, while the Lorentz ratio is
increased in the elastic transport. The reason of the difference
was investigated in figure 1. In high temperatures, the Lorentz
ratio decreases significantly because the electrical conductance
is increased. Such behavior was previously reported for the
single-level quantum dots~\cite{ref15}.
\begin{figure}[htb]
\begin{center}
\includegraphics[height=80mm,width=80mm,angle=0]{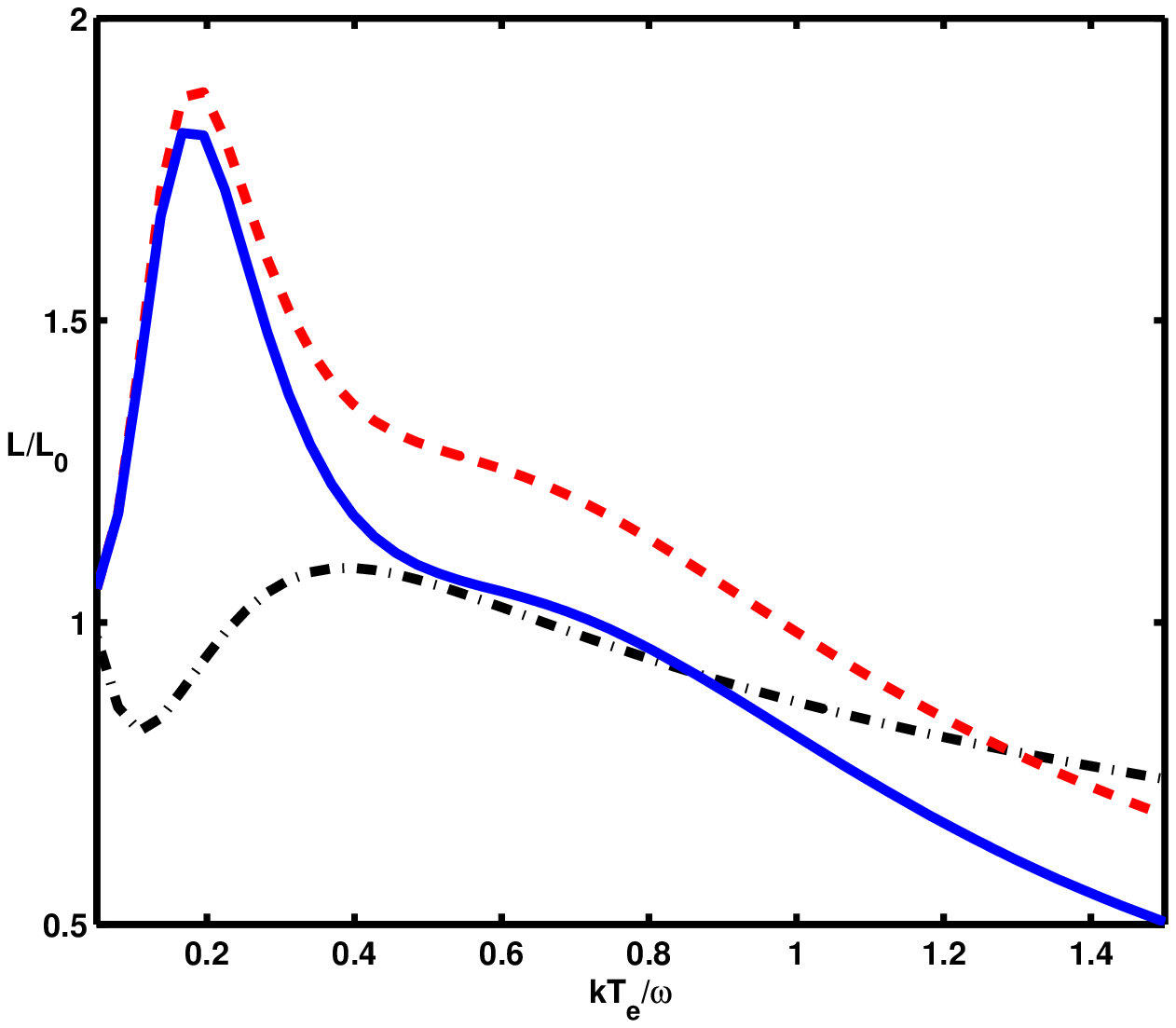}
\caption{ Dependence of Lorentz ratio on the electron
temperature.}\label{fig:6}       % Give a unique label
\end{center}
\end{figure}
\section{Summary}
\label{Summary} We analyze the thermoelectric properties of a
two-level molecule by means of equation of motion technique
within the  Green function formalism. The influence of the
Coulomb interactions, electron-phonon coupling, and temperature
on the figure of merit is examined. The electron temperature of
the bulky electrodes can be different from the phonon temperature
because of the smaller heat capacity of the phonons so that the
electron-phonon nonequilibrium thermodynamics may govern the
system. The temperature difference results in the novel and
interesting phenomena. It is observed that the electron-phonon
interaction results in the reduction of the number of peaks in
the thermal conductance due to the disturbance of the symmetry of
the DOS near the chemical potential of the leads. Furthermore,
the Coulomb repulsions and electron-phonon interaction result in
the violation of the Wiedemann-Franz law.

\end{document}